# Hyperfine Wave Functions and Force Densities for the Hydrogen Atom

F. J. Himpsel

Department of Physics, University of Wisconsin Madison,
1150 University Ave., Madison, WI 53706, USA, fhimpsel@wisc.edu

**Abstract**

This study addresses the effect of the magnetic hyperfine interaction on the relativistic H $1s$ wave functions. These are used to calculate the electric, magnetic, and confinement force densities acting on the $1s$ electron. The magnetic field couples Dirac equations for different angular momenta. These are solved numerically for the hyperfine singlet and triplet, as well as for a classical magnetic dipole. In the singlet ground state the hyperfine interaction shifts the electron density toward the proton. A similar shift is found for the classical dipole, and an opposite shift for the triplet. The cross-over between charge accumulation and depletion occurs at 1.325 times the Bohr radius. The behavior of the wave functions is investigated down to distances smaller than the proton radius, including the incorporation of virtual positrons. The force densities are determined and balanced against each other.

**Contents**

## 1. Introduction

This work is part of a program to investigate the internal stability of prototypical structures in atomic and particle physics. Two previous publications demonstrated a local force balance for the 1$s$ electron in the hydrogen atom [1] and for the vacuum polarization surrounding a point charge [2]. In both cases an exact balance between the attractive electrostatic force density and the repulsive confinement force density was derived from the Dirac equation. That goes beyond the common stability arguments for the H atom which stress a minimum in the total energy. In [1] the proton was approximated by a point charge without a magnetic moment, thereby keeping the forces isotropic. Here we ask whether the force densities remain balanced when the interaction with the magnetic dipole field of the proton is included.

The magnetic interaction gives rise to the hyperfine splitting of the energy levels which has been studied in great detail by perturbation theory [3]-[12]. It has impact in various areas of science. For example, the 21 cm radio line of atomic H corresponds to a transition between the singlet and triplet hyperfine levels of the 1$s$ electron. This line is used widely in astronomy for detecting hydrogen. High-precision measurements of fundamental constants often involve the H atom and its hyperfine levels. The poorly-known size and shape of the proton have limited the accuracy so far [4],[9]-[13]. Recently, this topic has come to the forefront after the proton radius obtained from muonic hydrogen [13] differed significantly from that obtained by electron scattering at normal hydrogen. This problem can be turned into an asset by using the hyperfine interaction of the H 1$s$ electron as sensitive probe of the magnetic dipole distribution inside the proton – an open question in quantum chromodynamics.

Driven by precision measurements the hyperfine energy levels of H have been studied in great detail. Perturbation theory is the technique of choice, because it takes advantage of the small hyperfine energy shifts. But in $2^{nd}$ order perturbation theory of the Dirac equation the integral over intermediate states actually diverges at large momenta [5],[6],[7]. Even in $1^{st}$ order the integrand of the radial matrix elements differs substantially between a point-like and extended nucleus at high Z (see Ref. [9], Fig. 1). Such difficulties at small distances suggest a non-perturbative solution of the Dirac equation for the hyperfine wave functions all the way down to the proton radius.

While the magnetic interaction is reduced by a factor of $10^{-3}\alpha$ from the Coulomb interaction at the Bohr radius of the 1$s$ electron ($a_0 = 1/\alpha m_e \approx 5 \cdot 10^{-11}$ m), it grows faster toward small distances. One encounters some characteristic length scales on the way to the proton radius. At the reduced Compton wavelength of the electron ($\lambdabar_C = 1/m_e \approx 4 \cdot 10^{-13}$ m) the interaction with virtual electrons and positrons begins to play a role, requiring quantum electrodynamics. Furthermore, the electric and magnetic potentials generated by the electron become comparable at ½$\lambdabar_C$. Next in line is the classical electron radius ($r_{class} = \alpha/m_e \approx 3 \cdot 10^{-15}$ m). Although irrelevant as such, it might have some other role in quantum electrodynamics. At the proton radius ($r_p \approx 1 \cdot 10^{-15}$ m) the strong interaction between the constituents of the proton comes into play.

This work is focused on wave functions and force densities rather than energy levels. The balance between all the force densities provides a detailed stability criterion for the H atom, and the wave function of the 1s electron is essential for calculating force densities. The Lorentz force density $\mathbf{f}=\rho\mathbf{E}+\mathbf{J}\times\mathbf{B}$ is obtained from the charge and current densities $\rho$ and $\mathbf{J}$ which in turn are determined by the electron wave function. The confinement force density is given by the divergence of the stress tensor which is again derived from the electron wave function (see [1] and references therein).

Before getting into detailed calculations it is helpful to discuss the effect of the proton's magnetic field on the various force densities at a qualitative level. Consider a proton with spin up ($m_p=+\frac{1}{2}$). The corresponding dipole moment $\boldsymbol{\mu}_p$ points up as well. In the singlet ground state the electron spin opposes the proton spin ($m=-\frac{1}{2}$), but its magnetic moment $\boldsymbol{\mu}_e$ is parallel to that of the proton. If one considers the proton as classical dipole, the magnetic energy $-\boldsymbol{\mu}_e\cdot\mathbf{B}_p$ is attractive along the z-axis and repulsive in the x,y-plane. Therefore one would expect the spherical probability density of the H 1s electron to become compressed along the z-axis and expanded in the x,y-plane. That would imply a non-vanishing electric quadrupole moment for the H atom in its ground state [15].

In reality, the proton does not behave like a classical magnetic dipole, since its dipole moment is generated by the quantum-mechanical spin operator $\mathbf{I}$ via the gyromagnetic ratio [3]-[14]. The magnetic energy is determined by the coupling of $\mathbf{I}$ with the angular momentum operator $\mathbf{J}$ of the 1s electron which yields the combined angular momentum operator $\mathbf{F}$. Since $\mathbf{F}$ vanishes for the singlet ground state, the charge distribution is isotropic – contrary to the classical picture.

In the following the angular and radial wave functions of the H 1s electron will be determined by solving the Dirac equation numerically for the classical and quantum-mechanical models of the magnetic field generated by the proton. The $1s_{1/2}$ wave function remains isotropic in the singlet ground state, since it couples only to other isotropic $s_{1/2}$ wave functions. The triplet state and the classical dipole exhibit additional couplings to anisotropic $d_{3/2}$ wave functions. Those introduce a small anisotropy in the charge density which corresponds to a finite electric quadrupole moment [15]. Here we consider exclusively the 1s electron which exhibits the highest charge density at small distances. That is where the magnetic forces becomes significant and the spin distribution of the proton comes into play. The 1s electron also represents the ground state of hydrogen where the force densities should be balanced.

**2. The Dirac Equation in a Magnetic Dipole Field**

Following [6],[14] we consider the interaction of the magnetic moment $\boldsymbol{\mu}_e$ of the electron with the magnetic field $\mathbf{B}$ and the vector potential $\mathbf{A}$ generated by the magnetic moment $\boldsymbol{\mu}_p$ of the proton. $\boldsymbol{\mu}_e$ and $\boldsymbol{\mu}_p$ are related to the angular momentum operators $\mathbf{J}$ and $\mathbf{I}$ of the electron and proton. (Bold face indicates three-vectors.) In contrast to most hyperfine calculations we are interested mainly in the wave functions, since those determine the force densities. Therefore we embark on a non-perturbative, numerical solution of the Dirac equation for the $1s_{1/2}$ electron in the H atom which includes both the electrostatic potential $\Phi$ and the magnetic dipole potential $\mathbf{A}$. Gaussian electromagnetic units combined with $\hbar,c=1$ lead to the following Dirac equation:

(1)  $[\gamma^\mu(i\partial_\mu - qA_\mu) - m_e]\psi = 0$    $A_\mu = (\Phi, -\mathbf{A})$    $q = q_e = -e$    $\alpha = e^2$

(2)  $q\gamma^\mu A_\mu = -e(\gamma^0 \Phi - \boldsymbol{\gamma} \cdot \mathbf{A}) = \gamma^0(H_{elec} + H_{mag})$

(3)  $\Phi = +e/r$    $\mathbf{A} = (\boldsymbol{\mu}_p \times \mathbf{e}_r) \cdot r^{-2}$    $r = |\mathbf{r}|$    $\mathbf{e}_r = \mathbf{r}/r$    $\mathbf{e}_z \times \mathbf{e}_r = \sin\theta\, \mathbf{e}_\varphi$

$H_{elec} = -\alpha/r$    $H_{mag} = e\,\gamma^0 \boldsymbol{\gamma} \cdot (\boldsymbol{\mu}_p \times \mathbf{e}_r) \cdot r^{-2}$

(4a)  $\boldsymbol{\mu}_p = \mu_p 2\mathbf{I}$    (quantum)    $H_{mag} \to H_{qu} = e\mu_p\,\gamma^0 \boldsymbol{\gamma} \cdot (2\mathbf{I} \times \mathbf{e}_r) \cdot r^{-2}$

(4b)  $\boldsymbol{\mu}_p = \mu_p \mathbf{e}_z$    (classical)    $H_{mag} \to H_{cl} = e\mu_p\,\gamma^0 \boldsymbol{\gamma} \cdot (\mathbf{e}_z \times \mathbf{e}_r) \cdot r^{-2}$

(5)  $\mu_p = b(e/2m_e)$    $b = 1.521\,032\,2... \cdot 10^{-3}$    $e\mu_p = \alpha(b/2m_e)$

Two options for the magnetic coupling are considered in (4a,b). The Hamiltonian $H_{qu}$ in (4a) uses spin operator $\mathbf{I}$ of the proton as quantum-mechanical source of the vector potential $\mathbf{A}$. This differs from the treatment of the Coulomb potential $\Phi$ which is derived from a classical point charge. Therefore we also explore the Hamiltonian $H_{cl}$ for a classical point dipole in (4b). $b$ is the magnetic moment $\mu_p$ of the proton in units of the Bohr magneton which contains the two basic units $e$ and $m_e$ in the Dirac equation. The g-factors of proton and electron are already incorporated in this definition.

To handle the strongly-varying magnetic field near r=0 the Dirac equation is solved without using perturbation theory. Nevertheless, we take advantage of the weak hyperfine coupling by extracting the Dirac equation for the pure Coulomb solution $\psi_C$ analytically and solving the remainder of the Dirac equation numerically. The magnetic Hamiltonian couples $\psi_C$ to wave functions $\psi_\kappa$ with compatible angular momentum quantum numbers of the form $\kappa = -1, +2, -3, +4, \ldots$ while the z-component $m_j = m$ and the parity $(-)^l$ remain unchanged [14]. The resulting ansatz for the total wave function $\psi$ of the H$1s$ electron is dominated by the Coulomb solution $\psi_C$ with $\kappa = -1$ and energy $E_C = (1-\alpha^2)^{1/2} \cdot m_e$ (compare Ref. [1], Eq. (15)). $\psi_C$ couples weakly to the remaining wave functions $\psi_{\kappa,m}$:

(6)  $\psi(\mathbf{r},t) = \psi_C(\mathbf{r}) + \Sigma_\kappa \psi_\kappa(\mathbf{r}) \cdot e^{-iEt}$    $\kappa = (-)^k \cdot k$    $k = 1, 2, 3 \ldots$

$m_p = +½$ for the proton    $m = -½$ for the singlet    $m = +½$ for the triplet

(7)  $\psi_C(\mathbf{r}) = \begin{pmatrix} g_C(r) \cdot \chi_{-1}^m(\theta,\varphi) \\ i\,f_C(r) \cdot \chi_{+1}^m(\theta,\varphi) \end{pmatrix}$    $\psi_\kappa(\mathbf{r}) = \begin{pmatrix} g_\kappa(r) \cdot \chi_\kappa^m(\theta,\varphi) \\ i\,f_\kappa(r) \cdot \chi_{-\kappa}^m(\theta,\varphi) \end{pmatrix}$

The functions $\chi_\kappa^m(\theta,\varphi)$ are standard spherical Pauli spinors satisfying $\int \chi_\kappa^{m\dagger} \chi_{\kappa'}^m\, d\Omega = \delta_{\kappa\kappa'}$. The quantum number $\kappa$ is the eigenvalue of the operator $-\gamma^0(\boldsymbol{\sigma} \cdot \mathbf{L} + 1)$ applied to the Dirac spinor $\psi$. The same eigenvalue $\kappa$ is obtained by applying the two-component operator $-(\boldsymbol{\tau} \cdot \mathbf{L} + 1)$ to the Pauli spinor $\chi_\kappa^m$. The vector $\boldsymbol{\tau} = (\tau^x, \tau^y, \tau^z)$ contains the Pauli matrices and the vector $\boldsymbol{\sigma}$ contains 4×4 matrices with two Pauli matrices along the diagonal. ½$\boldsymbol{\sigma}$ represents the spin operator $\mathbf{S}$ of the electron. It is also useful to define the spherical Dirac matrices $(\gamma^r, \gamma^\theta, \gamma^\varphi)$ which are composed of spherical Pauli matrices $(\tau^r, \tau^\theta, \tau^\varphi)$. Those are obtained by projecting the vector of Pauli matrices $\boldsymbol{\tau}$ onto the unit vectors $\mathbf{e}_r, \mathbf{e}_\theta, \mathbf{e}_\varphi$. The Hamiltonian of the classical magnetic dipole potential in (4b) contains the Dirac matrix $\gamma^\varphi$:

(8) $\gamma \cdot (\mathbf{e}_z \times \mathbf{e}_r) = \sin\theta\, \gamma \cdot \mathbf{e}_\varphi$ $\quad \gamma \cdot \mathbf{e}_\varphi = \gamma^\varphi = \begin{bmatrix} 0 & \tau^\varphi \\ -\tau^\varphi & 0 \end{bmatrix} \quad \tau^\varphi = \tau \cdot \mathbf{e}_\varphi = i \cdot \begin{bmatrix} 0 & -e^{-i\varphi} \\ e^{i\varphi} & 0 \end{bmatrix}$

Inserting the ansatz (6),(7) into the Dirac equation (2) and decomposing it into the orthonormal angular functions $\chi_\kappa^m$ yields a coupled set of radial Dirac equations for various angular momenta $\kappa$. For $\kappa=-1$ the Coulomb solution is extracted from the Dirac equation. This leads to three ranges of $|\kappa|$ with different Dirac equations:

(9a) $\partial_r F_C = +\frac{\kappa}{r} F_C + [m_e - E_C - \frac{\alpha}{r}] G_C$

$\partial_r G_C = -\frac{\kappa}{r} G_C + [m_e + E_C + \frac{\alpha}{r}] F_C$

$\partial_r F_\kappa = +\frac{\kappa}{r} F_\kappa + [m_e - E - \frac{\alpha}{r}] G_\kappa + \frac{b}{2m_e} \frac{\alpha}{r^2} [\Sigma_{\kappa'} a_{\kappa\kappa'} F_{\kappa'} + a_{\kappa,-1} \cdot F_C] - \delta E \cdot G_C$

$\partial_r G_\kappa = -\frac{\kappa}{r} G_\kappa + [m_e + E + \frac{\alpha}{r}] F_\kappa - \frac{b}{2m_e} \frac{\alpha}{r^2} [\Sigma_{\kappa'} a_{\kappa\kappa'} G_{\kappa'} + a_{\kappa,-1} \cdot G_C] + \delta E \cdot F_C$

$\left. \begin{array}{c} \psi_C \\ \end{array} \right\} \kappa = -1$

(9b) $\partial_r F_\kappa = +\frac{\kappa}{r} F_\kappa + [m_e - E - \frac{\alpha}{r}] G_\kappa + \frac{b}{2m_e} \frac{\alpha}{r^2} [\Sigma_{\kappa'} a_{\kappa\kappa'} F_{\kappa'} + a_{\kappa,-1} \cdot F_C]$

$\partial_r G_\kappa = -\frac{\kappa}{r} G_\kappa + [m_e + E + \frac{\alpha}{r}] F_\kappa - \frac{b}{2m_e} \frac{\alpha}{r^2} [\Sigma_{\kappa'} a_{\kappa\kappa'} G_{\kappa'} + a_{\kappa,-1} \cdot G_C]$

(9c) $\partial_r F_\kappa = +\frac{\kappa}{r} F_\kappa + [m_e - E - \frac{\alpha}{r}] G_\kappa + \frac{b}{2m_e} \frac{\alpha}{r^2} \cdot \Sigma_{\kappa'} a_{\kappa\kappa'} F_{\kappa'}$

$\partial_r G_\kappa = -\frac{\kappa}{r} G_\kappa + [m_e + E + \frac{\alpha}{r}] F_\kappa - \frac{b}{2m_e} \frac{\alpha}{r^2} \cdot \Sigma_{\kappa'} a_{\kappa\kappa'} G_{\kappa'}$

$\downarrow |\kappa|$

(9d) $F = r \cdot f \quad G = r \cdot g \quad E = E_C + \delta E \quad E_C = (1-\alpha^2)^{1/2} \cdot m_e$

Eq. (9a) is for $\kappa=-1$, (9b) for $|\kappa|>1$, $a_{\kappa,-1} \neq 0$, and (9c) for $a_{\kappa,-1}=0$. For $\kappa=-1$ there is an extra Dirac equation for the pure Coulomb solution $\psi_C$ which has an analytic solution. Since $\psi_C$ is $10^6$ times larger than $\psi_{-1}$, extracting it explicitly in (9a) greatly improves the precision of $\psi_{-1}$. Two remnants of $\psi_C$ remain in (9a) and (9b), one multiplied by the magnetic energy shift $\delta E \approx 10^{-7} \cdot E_C$ and the other by the equally small magnetic coupling $b\alpha/m_e \cdot r^{-2}$. The complete system (9a,b,c) represents a standard eigenvalue problem for $\delta E$.

The second $\kappa$-range in (9b) involves the wave functions $\psi_\kappa$ that couple to the Coulomb solution $\psi_C$ via non-zero coefficients $a_{\kappa,-1}$. That generates the term $a_{\kappa,-1} \cdot \psi_C$. For the highest range of $|\kappa|$ in (9c) the Dirac equations become homogeneous. In its generality this set of coupled Dirac equations looks rather imposing. It becomes fairly simple for the H1s singlet ground state where it is reduced to (9a). Even for the triplet and the classical dipole one needs to add only the contribution from $\kappa=+2$.

The coupling between different angular momenta is determined by the coefficients $a_{\kappa\kappa'}$. These are the angular matrix elements of the magnetic Hamiltonian $H_{mag}$ in (4a,b),(5). They interchange the upper and lower radial functions due to the vector $\gamma$ of anti-diagonal spatial Dirac matrices in $H_{mag}$. As a result, the $a_{\kappa\kappa'}$ become matrix elements between the upper and lower Pauli spinors of the Dirac spinor $\psi$ and vice versa, i.e., $\langle i\chi_{-\kappa'}^m | H_{mag} | \chi_\kappa^m \rangle$ and $\langle \chi_\kappa^m | H_{mag} | i\chi_{-\kappa'}^m \rangle$. Recombining them into Dirac spinors switches the upper and lower radial functions G and F. Non-vanishing matrix elements connect states with $\kappa=(-1)^k \cdot k$, i.e., even parity. The common $m$ is $-\frac{1}{2}$ for the singlet and $+\frac{1}{2}$ for the triplet. Conservation of the total angular momentum $\mathbf{F}=\mathbf{J}+\mathbf{I}$ requires $f'=f$, $m'_f=m_f$.

For the hyperfine interaction (4a) we convert the angular matrix elements published in Ref. [6], Eq. 12 from Wigner 6-j and 3-j symbols to a more explicit form, with the quantum numbers $\kappa$ and $\kappa'$ of the two angular wave functions as variables:

(10a) $a^{qu}_{\kappa\kappa'} = 2\cdot(-)^{j+i+f+\bar{l}'}\cdot[2(2j'+1)(2j+1)(2i+1)i(i+1)]^{1/2}\cdot\begin{Bmatrix} j' & i & f \\ i & j & 1 \end{Bmatrix}\cdot\begin{bmatrix} j' & j & 1 \\ \frac{1}{2} & \frac{1}{2} & -1 \end{bmatrix}$

$$= \begin{cases} \left. \begin{array}{ll} \frac{2|\kappa\kappa'|^{1/2}}{\kappa-\kappa'} & \text{for } |\kappa'|=|\kappa|-1,\ \kappa'=-\kappa+\text{sign}(\kappa) \\ \frac{-2\kappa}{2|\kappa|-1} & \text{for } \kappa'=\kappa \end{array} \right\} f=|\kappa|-1 \\ \left. \begin{array}{ll} \frac{2|\kappa\kappa'|^{1/2}}{\kappa'-\kappa} & \text{for } |\kappa'|=|\kappa|+1,\ \kappa'=-\kappa-\text{sign}(\kappa) \\ \frac{+2\kappa}{2|\kappa|+1} & \text{for } \kappa'=\kappa \end{array} \right\} f=|\kappa| \end{cases}$$

$a^{qu}_{\kappa\kappa'} \to 1$ for $|\kappa|\to\infty$

|  | $a^{qu}_{\kappa\kappa'}$ | $\kappa'\ (l_j)$ |  |
|---|---|---|---|
| $f=0$ (singlet) | $+2$ | $-1\ (s_{1/2})$ |  |
| $f=1$ (triplet) | $-2/3$ | $-1\ (s_{1/2})$ | $\kappa=-1\ (s_{1/2})$ |
|  | $+\sqrt{8}/3$ | $+2\ (d_{3/2})$ |  |
|  | $+\sqrt{8}/3$ | $-1\ (s_{1/2})$ |  |
|  | $-4/3$ | $+2\ (d_{3/2})$ | $\kappa=+2\ (d_{3/2})$ |
|  | $+4/5$ | $+2\ (d_{3/2})$ |  |
|  | $-\sqrt{24}/5$ | $-3\ (d_{5/2})$ |  |

$\kappa=(-)^{l+1}\cdot(j+\frac{1}{2})$
$j=|\kappa|-\frac{1}{2}$
$l=|+\kappa+\frac{1}{2}|-\frac{1}{2}$
$\bar{l}=|-\kappa+\frac{1}{2}|-\frac{1}{2}$

The $1s_{1/2}$ singlet state $\psi_C$ with $\kappa=-1, j=\frac{1}{2}, f=0$ couples only to wave functions $\psi_{-1}$ with $s_{1/2}$ angular symmetry, forming the two Dirac equations (9a). In all other cases the there are two non-vanishing matrix elements for each combination of $j$ and $f$. The $1s_{1/2}$ triplet state, for example, couples not only to the $s_{1/2}$ wave function ($\psi_C+\psi_{-1}$) but also to the $d_{3/2}$ wave function $\psi_{+2}$. The latter forms the Dirac equation (9b). These coupled Dirac equations form a self-contained system determining $\psi_{-1}$ and $\psi_{+2}$. A coupling to $d_{5/2}$ is avoided by setting $\psi_{-3}$ to zero. The $d_{3/2}$ and $d_{5/2}$ wave functions form a separate triplet with the $3d_{3/2}$ Coulomb solution.

For the classical magnetic field we calculate the angular matrix elements explicitly, using the Hamiltonian (4b) and the representation of the spinors $\chi^m_\kappa$ defined in the Appendix of [1].

(10b) $a^{cl}_{\kappa\kappa'} = \begin{cases} \frac{|\kappa\kappa'|^{1/2}}{\kappa-\kappa'} & \text{for } |\kappa'|=|\kappa|-1,\ \kappa'=-\kappa+\text{sign}(\kappa) \\ \frac{-2\kappa}{4\kappa^2-1} & \text{for } \kappa'=\kappa \\ \frac{|\kappa\kappa'|^{1/2}}{\kappa'-\kappa} & \text{for } |\kappa'|=|\kappa|+1,\ \kappa'=-\kappa-\text{sign}(\kappa) \end{cases}$

$$a_{\kappa\kappa'}^{cl} \to \begin{cases} \frac{1}{2}|\kappa|^{-1} & \text{for } \kappa' = \kappa \\ \frac{1}{2} & \text{for } |\kappa'| = |\kappa| \pm 1 \end{cases} \quad |\kappa| \to \infty$$

$$\text{bonding state} \begin{cases} \begin{array}{ll} a_{\kappa\kappa'}^{cl} & \kappa'\ (l_j) \\ +\frac{2}{3} & -1\ (s_{1/2}) \\ +\sqrt{2}/3 & +2\ (d_{3/2}) \end{array} \Bigg\} \kappa = -1\ (s_{1/2}) \\ \begin{array}{ll} +\sqrt{2}/3 & -1\ (s_{1/2}) \\ -\frac{4}{15} & +2\ (d_{3/2}) \\ -\sqrt{6}/5 & -3\ (d_{5/2}) \end{array} \Bigg\} \kappa = +2\ (d_{3/2}) \\ \begin{array}{ll} -\sqrt{6}/5 & +2\ (d_{3/2}) \\ +\frac{6}{35} & -3\ (d_{5/2}) \\ +\sqrt{12}/7 & +4\ (g_{7/2}) \end{array} \Bigg\} \kappa = -3\ (d_{5/2}) \end{cases}$$

Compared to the hyperfine matrix elements each $\kappa$ couples to an additional $\kappa'$, resulting in two possible $\kappa'$ values for $\kappa = -1$ and three for higher $|\kappa|$. For the lowest level one can find a solution containing just the $s_{1/2}$ and $d_{3/2}$ wave functions $\psi_C, \psi_{-1}$, and $\psi_{+2}$.

The differences between the classical and quantum dipoles can be traced to their angular momenta. While the hyperfine field is proportional to the nuclear spin operator **I** with angular momentum ½, the classical dipole field is proportional to the vector potential **A** whose angular momentum is 1. Half-integer angular momenta cannot be treated as classical fields. Treating both Hamiltonians quantum-mechanically provides a simple explanation for the different patterns of angular matrix elements. This is illustrated in Figure 1.

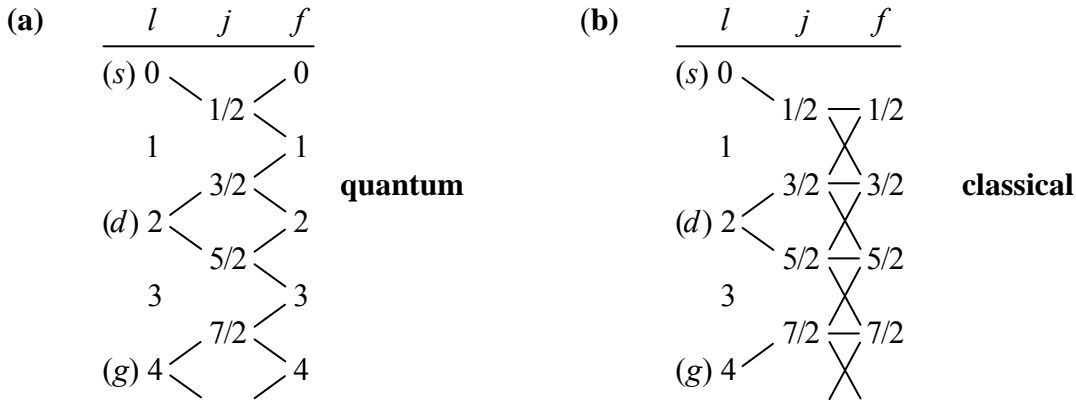

**Figure 1** Explanation of the angular couplings induced by (a) the hyperfine field and (b) the classical magnetic dipole field. Allowed **L·S** and **J·I** couplings are shown as lines, using the addition rules for angular momenta. The classical vector potential **A** is treated as an angular momentum **I_A** with quantum number $i_A = 1$. The coupling **J·I_A** gives rise to the total angular momentum **F**.

The hyperfine Hamiltonian $H_{qu}$ couples the angular momenta of electron and proton, $j = l \pm \frac{1}{2}$ and $i = \frac{1}{2}$, resulting in two possible combined angular momenta $f = j \pm \frac{1}{2}$. The angular momentum $i_A = 1$ assigned to the classical vector potential **A** produces three combined angular momenta $f_A = j, j\pm 1$. The $s_{1/2}$ state with $j=\tfrac{1}{2}$ has only two couplings, since $f_A = j-1$ would be negative. The states contributing to a particular energy level on the right are obtained by tracing the lines for the allowed couplings back to the left.

## 3. Solution of the Dirac Equation for the H 1s Electron

The singlet $1s_{1/2}$ hyperfine state with $\kappa=-1$ couples only to states with $\kappa'=-1$ via the matrix element $a^{qu}_{\kappa\kappa'}=+2$. That leads to the Dirac equations (9a) with $E, \delta E$ from (9d):

(11) $\quad \partial_r F_{-1} = -\tfrac{1}{r} F_{-1} + [m_e - E - \tfrac{\alpha}{r}] G_{-1} + \tfrac{b}{2m_e}\tfrac{\alpha}{r^2}\cdot 2(F_{-1} + F_C) - \delta E \cdot G_C$

$\quad\quad\quad \partial_r G_{-1} = +\tfrac{1}{r} G_{-1} + [m_e + E + \tfrac{\alpha}{r}] F_{-1} - \tfrac{b}{2m_e}\tfrac{\alpha}{r^2}\cdot 2(G_{-1} + G_C) + \delta E \cdot F_C$

The Dirac equations for the triplet and the the classical dipole are given in (A1). The strategy for solving them is described in Appendix A. The resulting radial wave functions are shown in Figures 2 and 3, and the angular wave functions are sketched in Figure 4.

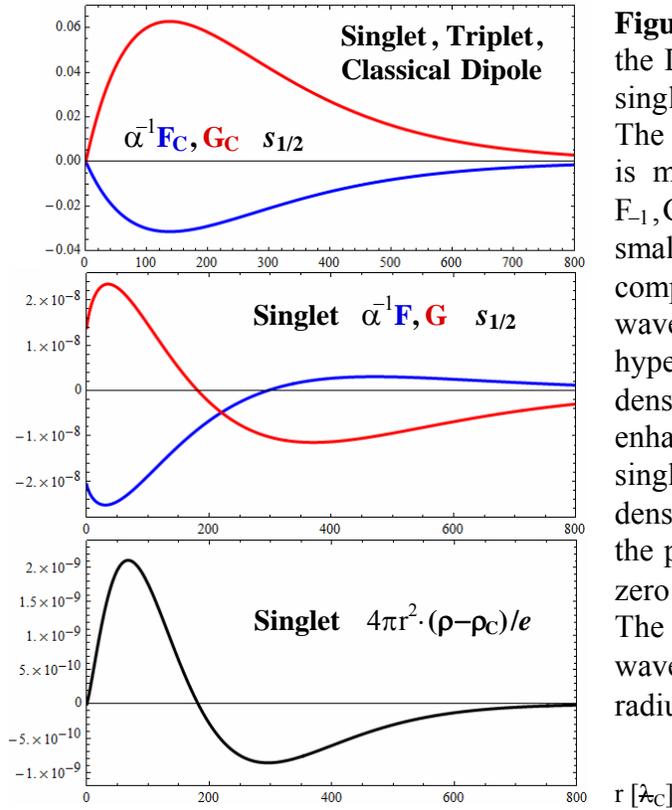

**Figure 2** Numerical solution of the Dirac equation (11) for the hyperfine singlet state forming the ground state of H. The pure Coulomb solution $F_C, G_C$ (top) is modified by the $s_{1/2}$ wave functions $F_{-1}, G_{-1}$ (center). The latter are $10^6$ times smaller. The sign change of the large component $G_{-1}$ ensures that the total wave function remains normalized. The hyperfine interaction shifts the electron density $\rho$ toward the proton and thereby enhances the bonding character of the singlet (bottom). The peak of the induced density lies at $r\approx 67\,\lambdabar_C \approx 0.5\,a_0$. Far from the proton the density is depleted, with a zero crossing at $r\approx 181.6\,\lambdabar_C \approx 1.325\,a_0$. The length unit is the reduced Compton wavelength $\lambdabar_C = 1/m_e$, which puts the Bohr radius at $a_0 = \alpha^{-1} \approx 137$.

The counterpart of Figure 2 for the triplet and classical dipole is Figure 3. The $s_{1/2}$ wave functions have similar shape, but are scaled by factors of $-\tfrac{1}{3}$ for the triplet and $+\tfrac{1}{3}$ for the classical dipole. These factors can be traced to different coefficients $a_{-1,-1}$ for coupling $F_{-1}, G_{-1}$ to the the Coulomb solution $F_C, G_C$ (and specifically to the ratio of the $a_{-1,-1}$). The same factors scale the hyperfine energy shifts $\delta E_1$ given in Eq. (13) below. For the antibonding triplet the electron density shifts away from the proton, and for the bonding ground state of the classical dipole it shifts toward the proton (compare the bottom panels of Fig. 3). Similarly, the $d_{3/2}$ wave functions differ by a factor of 2 between triplet and classical dipole. That is the ratio of their coupling coefficients $a_{+2,-1}$.

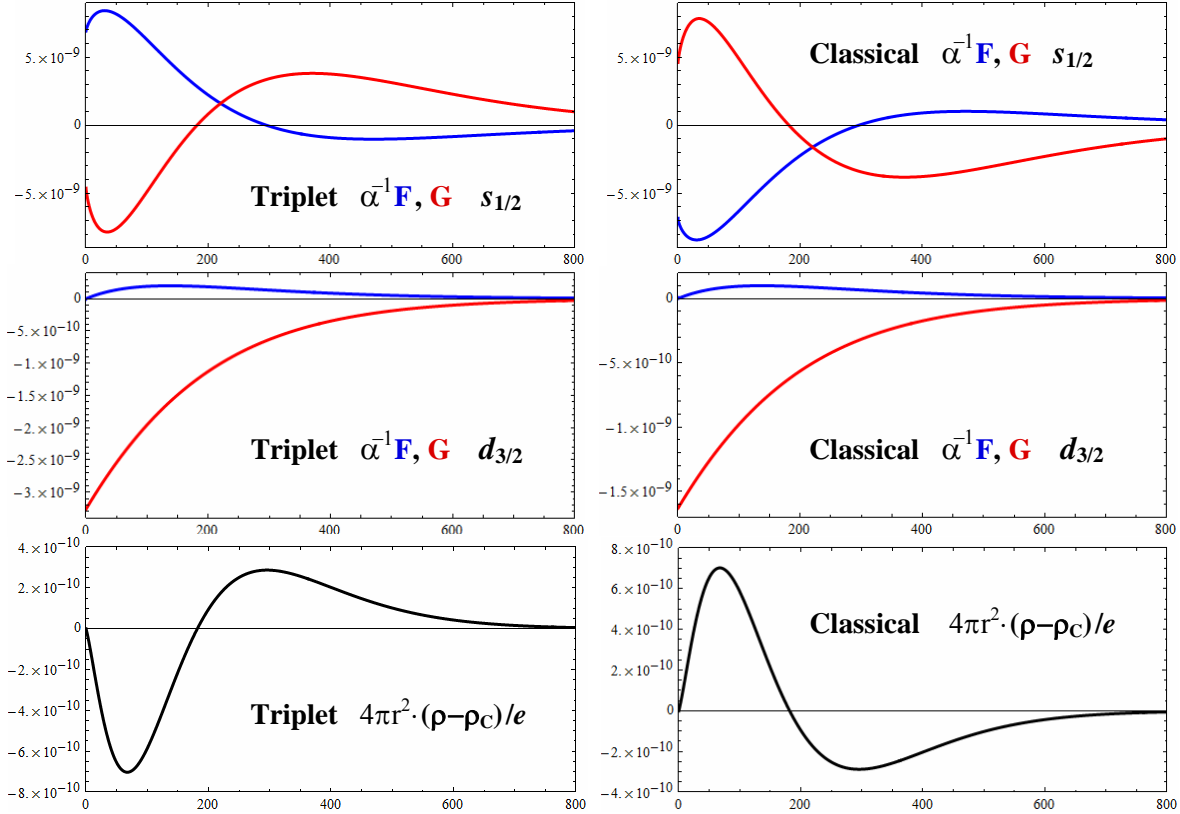

**Figure 3** Radial wave function analogous to Figure 2, but for the antibonding 1s triplet state and the bonding state of the classical point dipole. These are solutions of the Dirac equation given in (A1). The magnetic field generates the anisotropic $d_{3/2}$ wave functions $F_{+2}, G_{+2}$ in addition to the $s_{1/2}$ wave functions $F_{-1}, G_{-1}$.

The contributions to the energy eigenvalue E of the Dirac equation can be sorted by their order in $\alpha$. The energy $E_C$ of the pure Coulomb solution $\psi_C$ is of $O(\alpha^2)$. This holds for both the Coulomb energy and the opposing kinetic energy. The 1st order hyperfine energy $\delta E_1$ is of $O(\alpha^4)$, obtained from the expectation value of $H_{qu}$ with $\psi_C$ [14]. $\delta E_2$ is obtained as eigenvalue of the Dirac equation (11) and found to be of $O(\alpha^6)$:

(12) $\quad E = E_C + \delta E \qquad\qquad E_C = m_e \cdot \gamma \qquad \gamma = (1-\alpha^2)^{1/2}$

$\qquad \delta E = \delta E_1 + \delta E_2 \qquad \delta E_1 = \langle \psi_C | H_{qu} | \psi_C \rangle = \langle i\chi_{+1}^m | H_{qu} | \chi_{-1}^m \rangle \cdot \int 2 f_C g_C \, dr$

$\qquad \langle i\chi_{+1}^m | H_{qu} | \chi_{-1}^m \rangle = e\mu_p \cdot a_{-1,-1} \qquad \int 2 f_C g_C \, dr = -m_e^2 \cdot 2\alpha^3/[\gamma(2\gamma-1)] \qquad \mu_p = b \cdot (e/2m_e)$

(13) $\quad \delta E_1 = -m_e b \cdot 2\alpha^4/[\gamma(2\gamma-1)] \approx -m_e \dfrac{m_e}{M_p} g_p \cdot \alpha^4 \quad$ with $\quad g_e \approx 2, \gamma \approx 1, a_{-1,-1} = 2 \quad$ (singlet)

$\qquad \delta E_2 \approx 0.8 \, \alpha^2 \cdot \delta E_1$

$\delta E_1$ serves as reference for the contribution $\delta E_2$ from higher orders. Since $\delta E_1$ is negative, $\delta E_2$ shifts the singlet level down further. Such a downshift was also found by 2nd order perturbation theory [6].

The angular part of the wave functions is illustrated by the corresponding electron densities in Figure 4. The anisotropy of the hyperfine triplet and the classical dipole states is dominated by the interference term between $d_{3/2}$ and the $s_{1/2}$ Coulomb wave function (right side). The pure $d_{3/2}$ contribution at the center is much weaker.

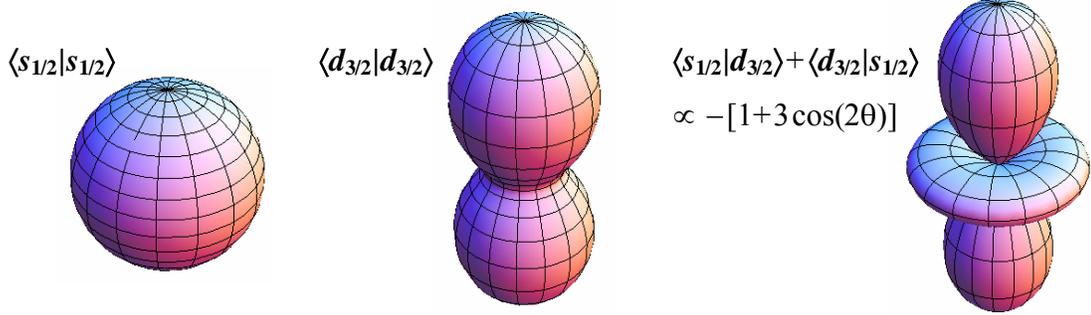

**Figure 4**  Angular wave functions of the hyperfine triplet and the classical dipole (for $m_j = ½$). In both cases the $s_{1/2}$ states (left) hybridize with $d_{3/2}$ states (center), creating a $s-d$ hybrid (right). Its polar and equatorial lobes have opposite signs. This anisotropic charge density leads to an electric dipole moment which is absent for the hyperfine singlet.

## 4. Electromagnetic Force Densities

With the wave functions in hand we can proceed to a calculation of the force densities that control the stability of the H atom. For the Coulomb solution $\psi_C$ the electrostatic and confinement force densities were found to balance each other exactly [1]. In the following we ask whether this remains true when the magnetic field of the proton is added to its electric field. The focus will be on the hyperfine singlet, since it represents the ground state of the H atom and therefore should have balanced force densities. Its wave function is isotropic due to the vanishing total angular momentum **F**. All force densities become radial. Following [1] we distinguish the electromagnetic force density $\mathbf{f}_{A\psi}$ and the confinement force density $\mathbf{f}_\psi$. The former originates from the interaction Lagrangian between the electromagnetic field $A_\mu$ and the Dirac field $\psi$ and the latter from the Dirac field itself. $\mathbf{f}_{A\psi}$ is obtained from the Lorentz force by replacing charge and current with the charge density $\rho$ and the current density **J**:

(14) $\quad \mathbf{f}_{A\psi} = \rho\mathbf{E} + \mathbf{J}\times\mathbf{B} = \mathbf{f}_E + \mathbf{f}_B \hfill \psi=(\psi_C+\psi_\kappa) \quad \kappa=-1$

$\rho = -e\cdot\psi^*\psi \quad = -e\cdot\frac{1}{4\pi}(f^2+g^2) \quad = -e\cdot(F^2+G^2)/4\pi r^2$

$\mathbf{J} = -e\cdot\psi^*\gamma^0\gamma\psi = +e\cdot\frac{1}{4\pi}(2fg)\cdot\sin\theta\cdot\mathbf{e}_\varphi = +e\cdot(2FG)/4\pi r^2 \cdot\sin\theta\cdot\mathbf{e}_\varphi$

$\psi_C$ is the Coulomb solution and $\psi_\kappa$ the hyperfine addition. The electrostatic part $\mathbf{f}_E$ of $\mathbf{f}_{A\psi}$ consists of contributions from the Coulomb and hyperfine solutions:

(15a) $\quad \mathbf{f}_{E,C} = \rho_C\mathbf{E} = -\alpha\cdot r^{-2}\cdot\psi_C^*\psi_C\cdot\mathbf{e}_r \hfill \mathbf{E}=e\,r^{-2}\cdot\mathbf{e}_r$

(15b) $\quad \mathbf{f}_{E,\kappa} = \rho_\kappa\mathbf{E} = -\alpha\cdot r^{-2}\cdot[(\psi_C^*\psi_\kappa+\psi_\kappa^*\psi_C)+\psi_\kappa^*\psi_\kappa]\cdot\mathbf{e}_r$

In terms of the radial functions one obtains:

(16a) $\quad 4\pi r^2\cdot\mathbf{f}_{E,C} = -\alpha\cdot r^{-2}\cdot(F_C^2+G_C^2)\cdot\mathbf{e}_r$

(16b) $\quad 4\pi r^2\cdot\mathbf{f}_{E,\kappa} = -\alpha\cdot r^{-2}[2(F_\kappa F_C+G_\kappa G_C)+(F_\kappa^2+G_\kappa^2)]\cdot\mathbf{e}_r \hfill \kappa=-1 \text{ for the singlet}$

These force densities are shown in Figure 5. To assess the influence of higher order force densities from quantum electrodynamics we also include the electrostatic force density $\mathbf{f}_{VP}$ acting on the vacuum polarization surrounding the proton from [2]:

(16c) $\quad 4\pi r^2 \cdot \mathbf{f}_{VP} = -\alpha \cdot \frac{8\alpha}{3\pi} r^2 \{ Ki_1(z) + z^{-1} \cdot K_0(z) - (1 - 2z^{-2}) \cdot K_1(z) \} \cdot \mathbf{e}_r \qquad z = 2m_e r \quad m_e = 1$

$$Ki_1(z) = \frac{\pi}{2} \{ 1 - z \cdot [K_0(z) \cdot \mathbf{L}_{-1}(z) + K_1(z) \cdot \mathbf{L}_0(z)] \}$$

It becomes comparable to the hyperfine forces near $8\lambda_C$ and reaches the Coulomb force near $3\lambda_C$. The electrostatic forces are all attractive at small r, with the hyperfine force $\mathbf{f}_{E,\kappa}$ switching its sign at the negative cusp.

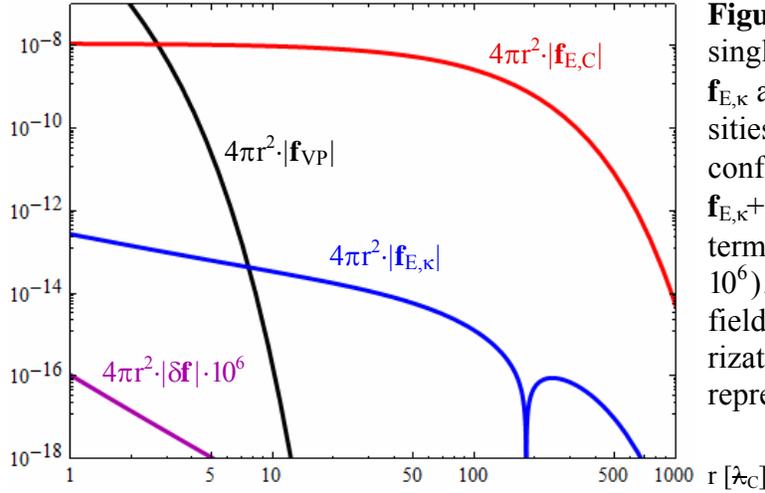

**Figure 5** Force densities in the singlet ground state of H. $\mathbf{f}_{E,C}$ and $\mathbf{f}_{E,\kappa}$ are the electrostatic force densities. $\mathbf{f}_{E,C}$ is compensated by the confinement force density $\mathbf{f}_{\psi,C}$, but $\mathbf{f}_{E,\kappa} + \mathbf{f}_{\psi,\kappa}$ leaves a tiny residual term $-\delta\mathbf{f}$ (amplified by a factor $10^6$). $\mathbf{f}_{VP}$ describes the Coulomb field acting on the vacuum polarization surrounding the proton. It represents radiative corrections.

The magnetic part $\mathbf{f}_B$ of $\mathbf{f}_{A\psi}$ is generated exclusively by the hyperfine interaction. It is well-defined only for the classical magnetic dipole field, using the vector potential $\mathbf{A} = \mu_p r^{-2} \sin\theta \cdot \mathbf{e}_\varphi$ from (3),(4b), $\mathbf{B} = \nabla \times \mathbf{A}$, and the hyperfine current density $\mathbf{J}_\kappa$:

(17) $\quad \mathbf{f}_{B,\kappa} = \mathbf{J}_\kappa \times \mathbf{B} \qquad \mathbf{J}_\kappa = +e \cdot [2(F_\kappa G_C + G_\kappa F_C) + 2F_\kappa G_\kappa]/4\pi r^2 \cdot \sin\theta \cdot \mathbf{e}_\varphi \qquad$ classical dipole

For the hyperfine interaction the situation becomes murky, since the Hamiltonion does not represent a classical magnetic field (see the discussion with Fig. 1). The main effect of the hyperfine interaction is a redistribution of the charge density which leads to an extra electrostatic force density $\mathbf{f}_{E,\kappa}$. It is almost balanced by the corresponding confinement force density $\mathbf{f}_{\psi,\kappa}$ except for a tiny residue $\delta\mathbf{f}$ (amplified by a factor $10^6$ in Fig. 5 to make it visible). This will be the topic of the next section.

## 5. Confinement Force Density and Force Balance

Before calculating the confinement force density some qualitative remarks are warranted. It was found in [1] that the electrostatic attraction is exactly balanced by the confinement repulsion for the pure Coulomb solution. It is tempting to make a similar argument for the additional force densities generated by the hyperfine interaction. The singlet exhibits an enhanced electron density near the proton which generates not only an additional Coulomb attraction but also increases the confinement repulsion. In the outer region the electron density is depleted which decreases both force densities. The turnaround at about $\frac{4}{3}a_0$ can be seen as a cusp of $\mathbf{f}_{E,\kappa}$ in Fig. 5.

The confinement force density is derived from the divergence of the stress tensor of the Dirac field $\psi$ (see Eq. (19) of Ref. [1]). For *s*-like wave functions one obtains:

(18) $\quad \mathbf{f}_\psi = -\frac{1}{4\pi}[2fg/r^2 + 2(fg'-f'g)/r + (fg''-f''g)] \cdot \mathbf{e}_r$

After making the substitutions $f \to F/r, g \to G/r$ and expanding the *s*-like singlet wave function $\psi = (\psi_C + \psi_\kappa)$ into its Coulomb and a hyperfine parts one obtains the following confinement force densities:

(19a) $\quad 4\pi r^2 \cdot \mathbf{f}_{\psi,C} = -[\quad 2F_C G_C/r^2 \quad + \quad (F_C G_C'' - F_C'' G_C) \quad] \cdot \mathbf{e}_r$

(19b) $\quad 4\pi r^2 \cdot \mathbf{f}_{\psi,\kappa} = -[\quad 2F_\kappa G_\kappa/r^2 \quad + \quad (F_\kappa G_\kappa'' - F_\kappa'' G_\kappa) \quad] \cdot \mathbf{e}_r$

$\quad\quad\quad\quad\quad\quad -[2(F_\kappa G_C + G_\kappa F_C)/r^2 + (F_\kappa G_C'' - F_\kappa'' G_C) + (F_C G_\kappa'' - F_C'' G_\kappa)] \cdot \mathbf{e}_r$

Since the confinement force density $\mathbf{f}_{\psi,C}$ exactly balances the electrostatic force density $\mathbf{f}_{E,C}$ for the Coulomb wave function, the question arises whether this remains true for the corresponding hyperfine force densities $\mathbf{f}_{\psi,\kappa}$ and $\mathbf{f}_{E,\kappa}$. In other words: are these four force densities balanced or is there a residual fifth (magnetic) force density?

The overall force density balance is tested analytically via the Dirac equation of the combined 1*s* wave function $\psi = (\psi_C + \psi_\kappa)$. For the singlet this yields a rather simple result, obtained by specializing (9c) to $\kappa = \kappa' = -1$ and setting $\psi_\kappa \to \psi$. This equation allows the elimination of the derivatives from the confinement force density $\mathbf{f}_\psi$ for comparison with the electrostatic force density $\mathbf{f}_E$ (using (16a),(19a) with $\psi_C \to \psi$ for $\mathbf{f}_E, \mathbf{f}_\psi$). The 2$^{nd}$ derivatives in (19a) are converted into 1$^{st}$ and 0$^{th}$ derivatives via the derivative of the Dirac equation. The 1$^{st}$ derivatives are then eliminated by the Dirac equation itself. Adding the two force densities of $\psi$ leaves a tiny residue $-\delta\mathbf{f}$ (see Fig. 5):

(20) $\quad 4\pi r^2 \cdot \delta\mathbf{f} = -4\pi r^2 \cdot (\mathbf{f}_E + \mathbf{f}_\psi)] = \frac{b\alpha}{m_e} r^{-3} \cdot 4FG$

The prefactor *b* shows that this residue originates from the hyperfine Hamiltonian. It also contains the product F·G which occurs in the current density **J** of the wave function $\psi$ in (14). That suggests a connection to the magnetic Lorentz force density $\mathbf{J} \times \mathbf{B}$ for classical fields. But the equivalent force density for the quantum-mechanical hyperfine field has yet to be established. By choosing the force density $\delta\mathbf{f}$ in (20) one could achieve a perfect force density balance for the H1*s* ground state, including the magnetic hyperfine interaction. This observation may help generalizing the canonical formalism established in classical field theory [1] to the quantum regime.

## 6. Summary

This study generalizes recent work on the stability of the H atom by adding the magnetic hyperfine interaction to the electrostatic Coulomb potential. In order to obtain a non-perturbative solution of the Dirac equation, a set of coupled Dirac equations is derived and solved numerically down to distances of less than $10^{-4} \lambda_C$. In addition to the singlet ground state of the 1*s* electron we also consider the triplet state and the ground state in a classical magnetic dipole field. The wave function remains isotropic for the singlet, despite the anisotropic magnetic dipole field of the proton. This is connected to the fact that the total angular momentum of proton and electron vanishes. The concept of a gyromagnetic ratio then predicts that the total magnetic moment vanishes. A more detailed look at the singlet spin wave function $(\uparrow_p\downarrow_e - \downarrow_p\uparrow_e)/\sqrt{2}$ reveals that the proton spin has equal probability of pointing up or down. Consequently the average magnetic field seen by the electron vanishes, thereby allowing an isotropic solution.

The main effect of the hyperfine interaction on the singlet wave function turns out to be electrostatic. Compared to the pure Coulomb solution, the electron density is compressed near the proton (and expanded on the outside). That explains why the singlet has bonding character. A similar effect is found for the ground state of a classical magnetic dipole field, but the binding energy is reduced by a factor of 3 because of the coefficients for angular momentum coupling. The hyperfine triplet has antibonding character with lower electron density near the proton. While the singlet remains isotropic, the triplet and the classical dipole become anisotropic due to an admixture of $d_{3/2}$ states. That leads to a finite electric quadrupole moment.

The force densities are evaluated for the singlet ground state. Since the magnetic hyperfine field is not a classical field, an automatic balance between electromagnetic and confinement force densities is not guaranteed. The canonical formalism for deriving force densities from the divergence of the stress tensor needs to be generalized from classical fields to quantum objects containing angular momentum operators or manybody quantum fields. The results obtained for the H atom will be helpful for assessing the stability of other fundamental objects in physics, such as muonic H [13], positronium [16], and possibly the electron itself [17].

**Appendix A: Numerical Solution of the Coupled Dirac Equations**

The coupled set of differential equations representing the Dirac equation for the hyperfine singlet haven been given already in (11). The corresponding set for the triplet and the classical dipole contains additional $d_{3/2}$ wave functions with $\kappa=+2$:

(A1)   Triplet:          $|\kappa|=1$: $a_{\kappa\kappa}=-\tfrac{2}{3}$    $|\kappa|=2$: $a_{\kappa\kappa}=-\tfrac{4}{3}$    $\kappa'\neq\kappa$: $a_{\kappa\kappa'}=a_{\kappa'\kappa}=+\sqrt{8}/3$

   Classical dipole:  $|\kappa|=1$: $a_{\kappa\kappa}=+\tfrac{2}{3}$    $|\kappa|=2$: $a_{\kappa\kappa}=-\tfrac{4}{15}$    $\kappa'\neq\kappa$: $a_{\kappa\kappa'}=a_{\kappa'\kappa}=+\sqrt{2}/3$

$$\kappa=-1 \begin{cases} \partial_r F_\kappa = +\tfrac{\kappa}{r} F_\kappa + [m_e-E-\tfrac{\alpha}{r}]G_\kappa + \tfrac{b}{2m_e}\tfrac{\alpha}{r^2}\left[a_{\kappa\kappa}(F_\kappa+F_C) + a_{\kappa\kappa'}F_{\kappa'}\right] - \delta E \cdot G_C \\ \partial_r G_\kappa = -\tfrac{\kappa}{r} G_\kappa + [m_e+E+\tfrac{\alpha}{r}]F_\kappa - \tfrac{b}{2m_e}\tfrac{\alpha}{r^2}\left[a_{\kappa\kappa}(G_\kappa+G_C) + a_{\kappa\kappa'}G_{\kappa'}\right] + \delta E \cdot F_C \end{cases}$$

$$\kappa=+2 \begin{cases} \partial_r F_\kappa = +\tfrac{\kappa}{r} F_\kappa + [m_e-E-\tfrac{\alpha}{r}]G_\kappa + \tfrac{b}{2m_e}\tfrac{\alpha}{r^2}\left[a_{\kappa\kappa}(F_\kappa+F_C) + a_{\kappa\kappa'}F_{\kappa'}\right] \\ \partial_r G_\kappa = -\tfrac{\kappa}{r} G_\kappa + [m_e+E+\tfrac{\alpha}{r}]F_\kappa - \tfrac{b}{2m_e}\tfrac{\alpha}{r^2}\left[a_{\kappa\kappa}(G_\kappa+G_C) + a_{\kappa\kappa'}G_{\kappa'}\right] \end{cases}$$

The coupling coefficients $a_{\kappa\kappa'}$ are obtained from (10a) for the triplet and from (10b) for the classical dipole. The numerical solution of the coupled Dirac equations is facilitated by the small magnitude of the magnetic interaction whose prefactor $e\mu_p = b/2m_e\cdot\alpha$ in (5),(11) contains both the electron/proton mass ratio and the fine structure constant $\alpha$. As a result the Dirac equations for different wave function become nearly decoupled. The dominant Coulomb solution $G_C, F_C$ is extracted at the outset by the ansatz (6) and handled as analytic inhomogeneous term in the remaining Dirac equations. That improves the numerical accuracy by focusing on the contributions $G_\kappa, F_\kappa$ from the magnetic interaction which are $10^6$ times smaller. But the standard eigenvalue problem of homogeneous Dirac equations becomes inhomogeneous, with $\psi_C$ generating two types of source terms in (A1). One is preceded by a coupling of the order $b \cdot \alpha/r^2$ and the other by $\delta E \propto b \cdot \alpha^4$.

For solving the coupled Dirac equations it is helpful to analyze constraints versus adjustable parameters. Each $\kappa$ generates a pair of 1$^{st}$ order Dirac equations for the radial functions $F_\kappa, G_\kappa$. Each pair requires two starting values which are adjusted such that both functions are regular at r=0 and r=∞. A single parameter is sufficient to tune both $F_\kappa$ and $G_\kappa$ at one boundary, since they are coupled so tightly that they vanish simultaneously. In addition there is a global constraint (the normalization of the combined wave function) and a global parameter (the common energy eigenvalue E). Strategies common to all fitting methods help speeding up the optimization, such as freezing unstable parameters while optimizing the rest and iterating this process with different parameter sub-sets.

The key to a quickly-converging solution is a strategy for finding the correct shape and amplitude of the dominant wave functions in a large parameter space. Once a sufficiently accurate approximation has been found by hand, the parameters can be optimized automatically to high precision. The large wave functions $G_\kappa$ need to be handled first. Their starting values need to be adjusted such that the solution becomes regular at r=0. Then the ratio $F_\kappa/G_\kappa$ can be varied to obtain a regular solution at r=∞. The wave function $G_{-1}$ is special, since it interferes with the Coulomb solution $\psi_C$. This interference term is the key to maintaining the normalization (see (A2) below). $\psi_{-1}$ must change its sign to leave two spatial regions with increased/diminished electron density that compensate each other in the normalization (compare Figs. 2,3).

Specifically, the coupled Dirac equations were solved from the starting point $r_{st}=800$ and integrated in both directions (down to $r_{min}=2 \cdot 10^{-5}$ and up to $r_{max}=1500$). A low starting point $r_{st}=10^{-1}$ was also used to investigate the asymptotics toward r=0. To handle the normalization a differential equation for the integral function of the electron density was solved in parallel with the Dirac equations. The *Mathematica* 9 routine *ParametricNDSolve* provided the numerical solution for a set of adjustable parameters, such as the energy eigenvalue E and the starting values of the radial wave functions. The starting value for E was the 1$^{st}$ order result $E_C + \delta E_1$ given in (12) for the singlet. The corresponding values for the triplet and the classical dipole were $-\frac{1}{3} \cdot \delta E_1$ and $+\frac{1}{3} \cdot \delta E_1$. Trial starting values for the wave functions were optimized by the routine *FindRoot* to satisfy the boundary conditions at r=0,∞ and the normalization.

An accurate normalization is obtained by subtracting the dominant term $\langle \psi_C | \psi_C \rangle = 1$ explicitly from the normalization of the total wave function $\psi$ in (6) and normalizing the rest to zero. The sum over $\kappa$ takes the orthogonality of the angular wave functions into account. Only $\psi_C$ and $\psi_{-1}$ interfere with each other:

(A2)     $1 = \langle \psi | \psi \rangle = \langle (\psi_C + \psi_{-1}) | (\psi_C + \psi_{-1}) \rangle + \Sigma_{\kappa \neq -1} \langle \psi_\kappa | \psi_\kappa \rangle$

        $0 = [\langle \psi_C | \psi_{-1} \rangle + \langle \psi_{-1} | \psi_C \rangle] + \Sigma_\kappa \langle \psi_\kappa | \psi_\kappa \rangle$

(A3)     $0 = \int \{ 2[F_{-1} F_C + G_{-1} G_C] + (F_{-1}^2 + G_{-1}^2) \} \, dr$            for the singlet

## Appendix B: Short Distance Behavior and the Role of Positrons

The non-perturbative solution of the Dirac equation at small distances opens a new avenue for testing strong magnetic interactions at small distances. Although this is not our primary goal, we have looked into the behavior of the singlet wave functions down to distances comparable to the proton radius. Above $r \approx 10^{-2} \lambda_C$ the wave functions $F_{-1}, G_{-1}$ converge toward well-defined limits at r=0, as shown in the left panel of Figure 6 for the singlet. The limits can be expressed in terms of $\alpha$ and $b$ from (5), with $m_e=1$:

(B1) $\quad \alpha^{-1} F_{-1}(r) \to -3 \cdot \alpha^{5/2} \cdot b(1+\frac{1}{2}b)$
$\quad\quad\quad G_{-1}(r) \to +2 \cdot \alpha^{5/2} \cdot b(1+\frac{1}{2}b)$ $\left.\right\} \quad \dfrac{F_{-1}(r)}{G_{-1}(r)} \to -\frac{3}{2}\alpha \quad$ for $r \to 0$

The scaling of the leading term is $\alpha^{5/2} \cdot b$, but that of the correction $\frac{1}{2}b$ is more complex.

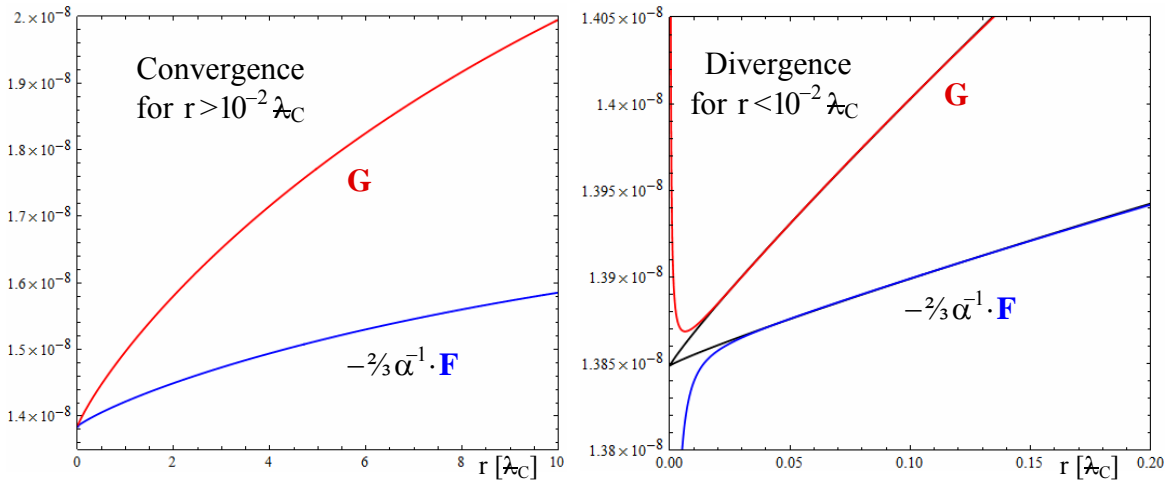

**Figure 6** Asymptotics of the singlet wave functions $F_{-1}, G_{-1}$ at small r. They appear to converge toward the analytic values given in (B1) for $r > 10^{-2} \lambda_C$ (left panel) but begin to diverge again below $10^{-2} \lambda_C$ (right panel). Extrapolation toward a finite common value at r=0 (black lines) yields optimized single-electron wave functions.

When analyzing the Dirac equation (11) for $r \to 0$ one realizes that the $1/r^2$ factor from the magnetic interaction needs to be matched by the derivative of a wave function similar to $\exp[\pm b\alpha/r]$. This function has an essential singularity at r=0, but the small prefactor $b\alpha \approx 10^{-5}$ keeps the function close to 1 for most of the r-range investigated here. Such behavior might be connected to the weak singularity found at large momenta in 2nd order perturbation theory [6]. A cure of this divergence will likely involve quantum electrodynamics, since electron-positron pairs are readily produced at those distances. Thus one could argue that the extrapolated wave functions in Fig. 6 provide the most accurate description of the hyperfine interaction within the limitations of the single-particle picture.

To judge the strength of the hyperfine corrections at various distances one can compare the prefactors of the electric and magnetic Hamiltonians in the Dirac equation (2). The two interactions are matched at the following distance:

(B2) $\quad \alpha/r = e\mu_p/r^2 = \alpha \cdot \frac{1}{2}b/r^2 \quad \Rightarrow \quad r \approx \frac{1}{2}b \cdot \lambda_C \approx 3 \cdot 10^{-16}$ m

This lies well inside the proton where the strong interaction takes over.

If one replaces the proton's magnetic moment $\mu_p$ by that of the electron ($\mu_e \approx \mu_p/b$), the small factor $b$ drops out. The two Hamiltonians are now equal at $½\lambdabar_C$:

(B3) $\quad e/r = \mu_e/r^2 \approx ½e/r^2 \quad \Rightarrow \quad r \approx ½\lambdabar_C \approx 2\cdot10^{-13}$ m

That describes positronium. A first approximation of the energy levels in positronium can be obtained from the hydrogenic model by introducing the relative cordinate ($\mathbf{r}_e - \mathbf{r}_p$), the reduced mass $\mu = ½m_e$, and the reduced Bohr radius $\mu/\alpha = 2a_0$ [16]. But going down to $½\lambdabar_C$ requires at least a fully-relativistic, two-particle Dirac equation, such as the Bethe-Salpeter equation and its modifications [18]. In addition one has to consider quantum electrodynamics which adds an extra Feynman diagram for the annihilation channel. Fortunately, that does not affect the singlet ground state, encouraging the solution of two-particle Dirac equations for positronium [18].

Considering distances below $½\lambdabar_C = 1/2m_e$ involves momenta larger than $2m_e$ which allow the creation of electron-positron pairs. As a result, even the two-particle picture becomes obsolete. It has been a long-standing question whether positrons can be combine with electrons in the Dirac equation without invoking full-fledged quantum electrodynamics. Positrons are related to the negative energy solutions of the Dirac equation which lead to the Klein paradox and to "Zitterbewegung" when mixed with positive energy solutions. An incorporation of positrons into the multi-electron Hartree-Fock-Dirac equations is more benign, but it hampers a variational solution due to "variational collapse" and "continuum dissolution" [19].

Technically, the wave functions for electrons with negative energy can be incorporated into the set of hyperfine Dirac equations. In order to share a common (positive) energy eigenvalue E they need to be converted to positrons with positive energy via charge conjugation [6]. During this process the angular momentum quantum numbers $\kappa$ and $m$ change their sign. In particular, an $s_{1/2}$ electron with the quantum numbers $-E, \kappa=-1, m=-½$ is transformed into a $p_{1/2}$ positron with quantum numbers $+E, \kappa=+1, m=+½$. Since the angular wave functions remain the same, such a positron can couple to the H$1s$ electron with coupling coefficients similar to those for the electron in (10a). The resulting extended set of coupled Dirac equations has been solved during the course of this work for the three cases in Figures 2, 3.

There has been a major problem, though. It is unclear how the electron density should be calculated in the presence of positron wave functions. The product of two positron wave functions clearly has to produce a negative electron density. But a product between a positron and an electron wave function does not have a well-defined sign. Such interference terms are possible, for example between the two states mentioned above. Omitting mixed products yields electron wave functions almost identical to those without positrons, since the positrons are coupled only weakly to the Coulomb solution.

Furthermore, one obtains positron wave functions that turn on already at the Bohr radius, while virtual $e^-e^+$ pairs should appear below $½\lambdabar_C$. For these reasons we have omitted positron wave functions, effectively adopting the "no pair" constraint from relativistic Hartree-Fock-Dirac theory [19]. An incorporation of positrons via relativistic two-particle Dirac equations might be option [18].